# An Algebraic Specification of the Semantic Web

Katerina Ksystra[1], Nikolaos Triantafyllou[1], Petros Stefaneas[2] and Panayiotis Frangos[1]

[1] School of Electrical and Computer Engineering, National Technical University of Athens,
Heroon Polytechniou 9, 15780 Zografou, Athens, Greece
{katksy, nitriant, pfrangos}@central.ntua.gr
[2] School of Applied Mathematical and Physical Sciences, National Technical University of
Athens, Heroon Polytechniou 9, 15780 Zografou, Athens, Greece
petros@math.ntua.gr

**Abstract.** We present a formal specification of the Semantic Web, as an extension of the World Wide Web using the well known algebraic specification language CafeOBJ. Our approach allows the description of the key elements of the Semantic Web technologies, in order to give a better understanding of the system, without getting involved with their implementation details that might not yet be standardized. This specification is part of our work in progress concerning the modeling the Social Semantic Web.

## 1 Introduction

As the Internet and the Web technologies are evolving, the need for specifying and formally describing them becomes more vital. In [9] Paine presents an algebraic specification of one particular kind of structured website, a slideshow, where each page is linked to its successor and predecessor. It uses CafeOBJ's module system to break this down into modules. This specification can be divided in two main parts. The first part, is a specification of the World Wide Web and the second defines this special site. In this paper we will make use of the first part only.

As it is known, the Semantic Web does not replace the current Web, they coexist. More precisely, the first supplements the second by adding semantics, meaning and new operations. For this reason, we have used the specification of the Web from [9] and by specifying the additional semantic technologies we obtained a specification of the Semantic Web.

We believe that by using formal methods to model the Semantic Web, we gain at least a better understanding of the underlying system. We also focus on how different parts work together, prove desired properties and study them within a general framework.

The rest of this paper is organized as follows: section 2 presents the specification of the Semantic Web as well as a brief introduction to CafeOBJ. In section 3 we analyze some future goals and thoughts and we conclude our paper.

## 2     Specification of the Semantic Web

We have extended the specification of [9] by adding various modules, in order to describe the fundamental components of the Semantic Web and some of their basic interactions. This was achieved with the use of CafeOBJ and thus before continuing we will give a short overview of the language.

### 2.1     CafeOBJ specification language

CafeOBJ is a language for writing formal specifications of models and verifying properties of them [1, 2]. It can be used to specify abstract data types as well as abstract state machines. A CafeOBJ specification consists of modules. In the body of the module we can declare the following: sorts, operators, variables and equations. A sort is a CafeOBJ term that describes abstract data types with multiple inheritance and operational semantics based on term rewriting [6, 7]. Declarations of operators start with op or ops if there are many. Operators are defined using equations; such a declaration starts with the keyword *eq* and the declaration of conditional equations with *ceq*. Also, the keyword used to denote a variable is *var*. The CafeOBJ system rewrites equations by regarding them as left-to-right rewrite rules.

### 2.2     Specification

One of the most important elements of CafeOBJ system is the module, as we said before. Thus, in figure 1 we show some of the modules we used as well as their connection. Two modules are connected when one imports another. We have used arrows that point to the module that is imported, to denote this relationship. As can be seen in figure 1, the module SEMANTICPAGE is linked to all the main modules of the specification (the grey-colored) and through this all the others are combined as well. Also, it is worth to mention that the module PAGES basically merges Paine's Specification with ours and thus we get a specification of the Semantic Web.

Below we also analyze two of the modules used in our specification so that it can be better understood how we can specify and express a concept of interest in CafeOBJ. These modules correspond to two building blocks of the RDF language, one of the main components of the Semantic Web. The first module corresponds to an RDF triple and the second to an RDF graph. But before explaining the modules we shall give a short description to these Semantic Web components.

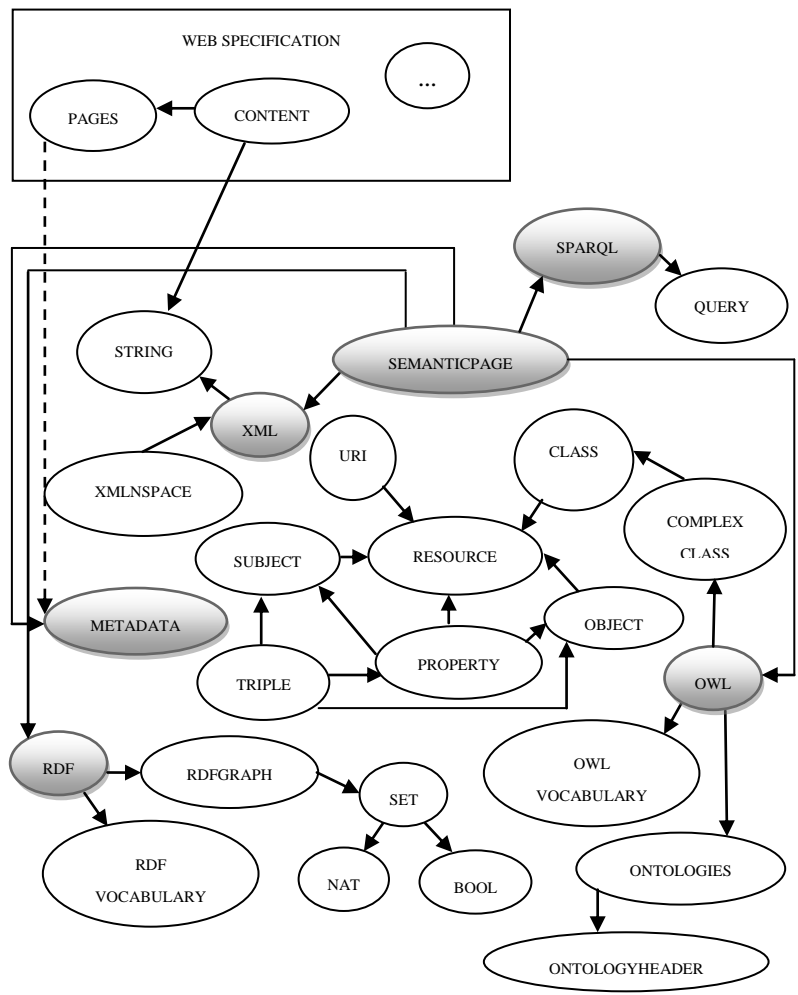

Fig. 1. Modules and their interaction

According to [10], the underlying structure of any expression in RDF is a collection of triples, each consisting of *a subject, a predicate and an object*. A set of such triples is called an RDF graph. This can be illustrated by a node and directed-arc diagram, in which each triple is represented as a node-arc-node link (hence the term graph). Each triple represents a statement of a relationship between the things denoted by the nodes that it links. The direction of the arc is significant: it always points toward the object. The following figure represents an RDF triple.

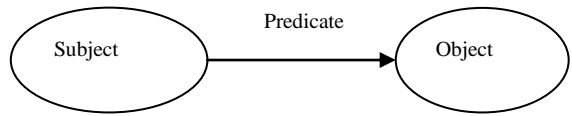

Fig. 2. RDF triple

The above can be captured in a CafeOBJ module as:

1. `mod! TRIPLE {`

2. `pr(SUBJECT)`

3. `pr(OBJECT1)`

4. `pr(PREDICATE)`

5. `[triple]`

6. `op rdftriple : subject predicate object -> triple}`

The first line simply denotes that this is a module with tight semantics named TRIPLE, while the fifth line states that the name of the sort that this module defines will be `triple` (note that it is not necessary to use the same name for the module and its sort). Lines 2, 3 and 4 import some modules, by protecting them, which are predefined and needed for the declaration of the operators that will construct the sentences of the module. Finally, in the last line we define an operator, the constructor of the sort triple, `rdftriple` that takes as arguments the sorts `subject`, `predicate` and `object` (as declared in the imported modules) and returns the sort `triple`. Let us see a second module that is a little more complex.

```
mod* RDFGRAPH {
pr(SET(TRIPLE {sort Elt -> triple})
*{sort Set -> rdfgraph, op empty -> empttr})}
```

This module is interesting because it is a view of a parameterized module. Here, we firstly import the parameterized module `SET`. This describes an arbitrary set whose elements are called Elt, a sort denoting an arbitrary element. We also ask from CafeOBJ to create a module with the name `RDFGRAPH`. This will also define a set but instead of arbitrary elements it will have `triple` as elements. Finally, the sort of this module will be `rdfgraph` and the constant that describes an empty graph will be called `empttr`. This specification captures the essence of an rdf graph as defined above.

Applying the same methodology presented here and having as a target to describe realistically the technologies of the Semantic Web we have created the rest of the modules of our specification. In the left column of the following table, we show some of the modules that were included in the specification of the Web while in the right side there are some of the modules that we constructed in order to specify the Semantic Web.

**Table 1.** Modules of the Web and the Semantic Web specification

| Web modules | Semantic Web modules |
|---|---|
| `Pages` | `Metadata` |
| `Content` | `Xml` |
| `Content'` | `Uri` |
| `Urls` | `Triple` |
| `Filenames` | `Rdfgraph` |
| `Html-utilities` | `Ontologies` |
| `Page-vs-Url` | `Owl` |
| `Url-vs-Filename` | `Resource` |
| `Document-Root-Mapping` | `Class` |
| | `Property` |
| | `Sparql` |
| | `Semanticpage` |

Apart from creating new modules that describe the Semantic Web we extended some of the existing modules by adding new operators. For example in the module `Pages` we used the operator `met` that takes as input a web page and returns its metadata. Some other operators can be seen in the table bellow:

**Table 2.** Operators of the Semantic Web specification

| | Operator's name | Input datatypes | Output datatypes |
|---|---|---|---|
| 1 | `op sameas        :` | `individual individual ->` | `owlstatement` |
| 2 | `op subpropertyof :` | `property property       ->` | `rdfstatement` |
| 3 | `op unionOf       :` | `class class             ->` | `complexclass` |
| 4 | `op merged?       :` | `uri uri                 ->` | `Bool` |

The first operator can be used to declare in OWL that two individuals are the same, while the second says in RDF terms that one property is a subproperty of another. The third operator takes two classes and creates a complex class from the union of them. The last operator takes as input two URIs and returns true or false if the URIs are merged or not respectively. Two URIs can be merged if they are identical and this is defined by the following equation:

```
ceq (merged?(U1,U2)) = true if (U1 = U2) .
```

## 3  Future work and conclusions

We work towards an abstract specification of the Semantic Web with the help of algebraic specifications. The choice of the level of abstraction was made according to the initial specification of the Web, so that the two specifications will be as coherent as possible. We believe that this first modeling can help to better comprehend how the components of the semantic web are combined and operate. Also, with this specification as a basis we can verify properties for the Semantic Web with the help of the OTS/CafeOBJ method [3, 4].

Our work can be used to specify the Social Semantic Web. In [7] it is stated that "the Social Semantic Web combines technologies, strategies and methodologies from the Semantic Web, social software and the Web 2.0". It seems natural, the specification of the Social Semantic Web to be a combination of the specifications of these three systems. The means of formally combining these entities will be given by the protocol composition method proposed in [8].